\documentclass[]{spie}  

\usepackage{amsmath,amsfonts,amssymb}
\usepackage{subcaption}
\usepackage{graphicx}
\usepackage{cite} 
\usepackage{epsfig}
\usepackage{amsmath}
\usepackage{nccmath}
\usepackage{amssymb}
\usepackage{mwe}
\usepackage{acro}
\usepackage{amssymb}
\usepackage{xcolor,colortbl}
\usepackage{tabularx}
\usepackage{relsize}
\usepackage{pifont}
\usepackage{booktabs} 
\usepackage{multirow}
\usepackage{multicol}
\usepackage{adjustbox}
\usepackage{float}
\usepackage{graphicx}
\usepackage{makecell}
\usepackage{tabu}
\usepackage[colorlinks=true, allcolors=blue]{hyperref}
\usepackage[capitalize]{cleveref}

\title{Assessment of Cell Nuclei AI Foundation Models in Kidney Pathology}

\author[a]{Junlin Guo}
\author[a]{Siqi Lu}
\author[b]{Can Cui}
\author[b]{Ruining Deng}
\author[b]{Tianyuan Yao}
\author[b]{Zhewen Tao}
\author[c]{Yizhe Lin}
\author[b]{Marilyn Lionts}
\author[b]{Quan Liu}
\author[a]{Juming Xiong}
\author[d]{Yu Wang}
\author[d]{Shilin Zhao}
\author[a,b]{Catie Chang}
\author[a]{Mitchell Wilkes}
\author[e]{Mengmeng Yin}
\author[e]{Haichun Yang}
\author[a,b,e]{Yuankai Huo}

\affil[a]{Department of Electrical and Computer Engineering, Vanderbilt University, Nashville, TN, USA}
\affil[b]{Department of Computer Science, Vanderbilt University, Nashville, TN, USA}
\affil[c]{Department of Mathematics, Vanderbilt University, Nashville, TN, USA}
\affil[d]{Department of Biostatistics, Vanderbilt University Medical Center, Nashville, TN, USA}
\affil[e]{Department of Pathology, Microbiology and Immunology, Vanderbilt University Medical Center, Nashville, TN, USA}

\authorinfo{Corresponding author: Yuankai Huo: E-mail: yuankai.huo@vanderbilt.edu}

\begin{document} 
\maketitle

\begin{abstract} 
Cell nuclei instance segmentation is a crucial task in digital kidney pathology. Traditional automatic segmentation methods often lack generalizability when applied to unseen datasets. Recently, the success of foundation models (FMs) has provided a more generalizable solution, potentially enabling the segmentation of any cell type. In this study, we perform a large-scale evaluation of three widely used state-of-the-art (SOTA) cell nuclei foundation models—Cellpose, StarDist, and CellViT. Specifically, we created a highly diverse evaluation dataset consisting of 2,542 kidney whole slide images (WSIs) collected from both human and rodent sources, encompassing various tissue types, sizes, and staining methods. To our knowledge, this is the largest-scale evaluation of its kind to date. Our quantitative analysis of the prediction distribution reveals a persistent performance gap in kidney pathology. Among the evaluated models, CellViT demonstrated superior performance in segmenting nuclei in kidney pathology. However, none of the foundation models are perfect; a performance gap remains in general nuclei segmentation for kidney pathology.

\end{abstract}

\keywords{Foundation model, Kidney Pathology, Instance Segmentation, Nuclei, Model Agreement, Ensemble learning}

\section{INTRODUCTION}
\label{sec:intro}  
Among the various tasks in digital pathology, cell nuclei instance segmentation is crucial as it is often the first step for extracting biological signals for accurate disease diagnosis and treatment planning~\cite{image_based_cell_profiling, deep_learning_pathology_survey, whole_cell_segmentation, structured_tumor_immune, medical_image_analysis_survey, image_based_cell_phenotyping}. However, a limitation in current methodologies is the generalizability of models in handling the growing scale and diversity of data. This challenge underscores the need for cell nuclei foundation models capable of segmenting nuclei from diverse organs and large-scale datasets~\cite{mediar, cellposs, cellvit, stardist}. Thus, we pose the question: how do current cell nuclei foundation models perform for comprehensive nuclei segmentation in kidney pathology? 

\begin{figure} [ht]
\begin{center}
\begin{tabular}{c} 
\includegraphics[width=1\linewidth]{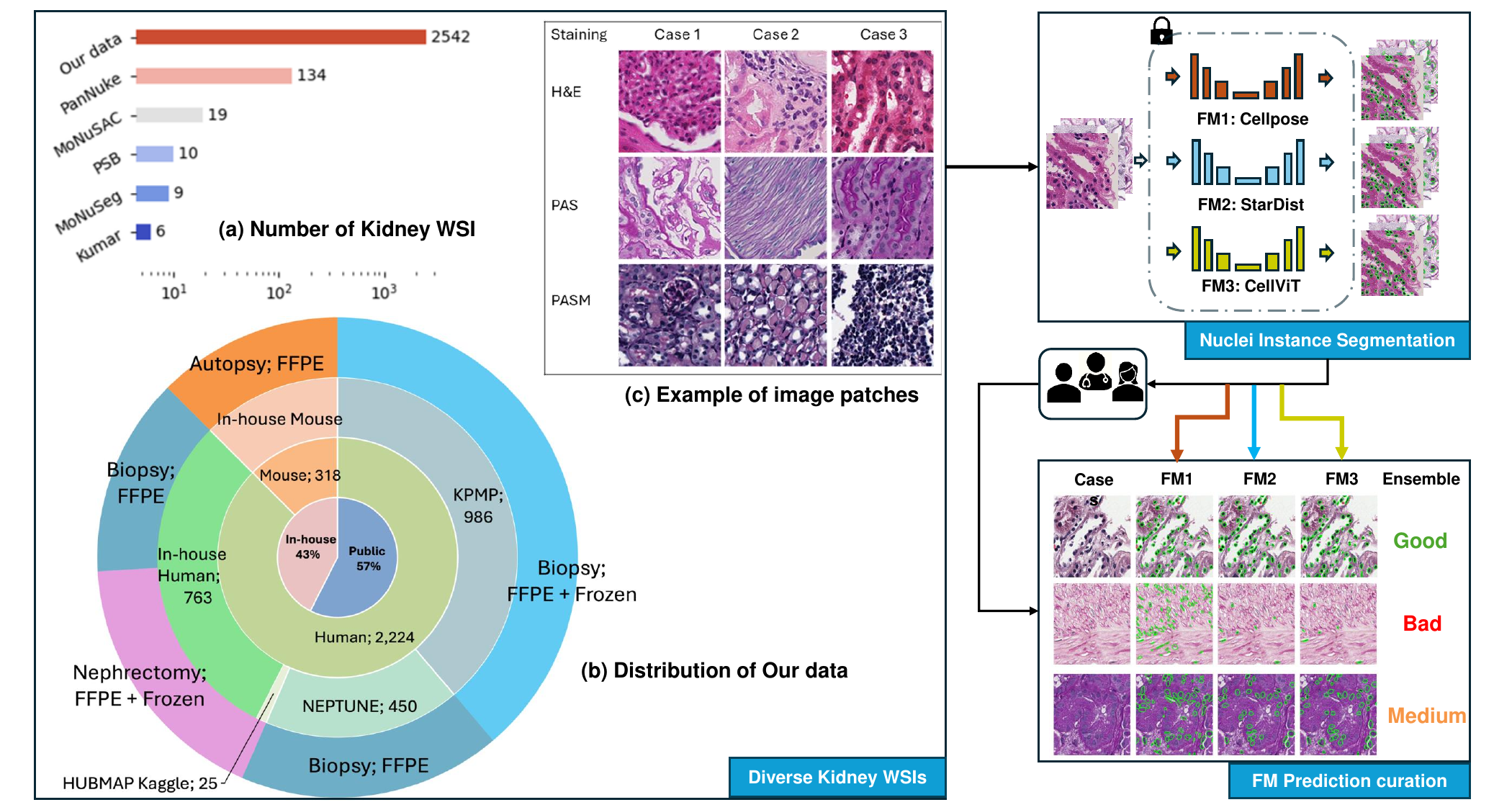}
\end{tabular}
\end{center}
\caption{Overall framework. (1) The creation of a diverse, large-scale kidney nuclei dataset.(2) Nuclei instance segmentation with Cellpose~\cite{cellposs}, StarDist~\cite{stardist}, and CellViT~\cite{cellvit}. (3) Rating on Segmentation Performance. Meaningful images were then identified through model agreement analysis.}
{ \label{fig:fig1}}
\end{figure}

To answer the preceding question, we evaluated the performance of current cell nuclei foundation models in kidney pathology. As shown in Fig.~\ref{fig:fig1}, we created a diverse evaluation dataset that includes kidney nuclei data from 2,542 kidney WSIs sourced from humans and rodents across both public and in-house datasets. The dataset includes various tissue types and sizes, such as biopsy, nephrectomy and autopsy samples, and features both formalin-fixed paraffin-embedded (FFPE) and frozen tissues. To our knowledge, the scale of this study's WSIs surpasses all publicly available labeled nuclei datasets that include the kidney, as illustrated in Fig.~\ref{fig:fig1}a. This work conducts a comparative analysis of three widely used SOTA cell nuclei foundation models --- Cellpose\cite{cellposs}, StarDist~\cite{stardist}, and CellViT~\cite{cellvit} --- on the task of nuclei instance segmentation. For validation, we include a manual rating-based system to curate foundation model predictions and identify failure samples across different models and modalities. To sum up, this study provides a comprehensive assessment of cell nuclei AI foundation models in the field of kidney pathology. The contribution of this paper is three-fold:

$\bullet$ \textbf{Creation of a Diverse, Large-Scale Kidney Nuclei Dataset:} The kidney dataset for evaluation consists of 2,542 kidney WSIs collected from both human and rodent sources, encompassing various tissue types, sizes, and staining methods. To our knowledge, this is the largest-scale evaluation of its kind to date. 

$\bullet$ \textbf{Rating-based segmentation curation with Foundation Models:} We utilized three SOTA foundation models (Cellpose\cite{cellposs}, StarDist~\cite{stardist}, and CellViT~\cite{cellvit}) in conjunction with a rating-based system to curate foundation model predictions. Then the evaluation can be shaped as analyzing distribution of the model predictions across the dataset. 

$\bullet$ \textbf{Agreement Analysis between Foundation Models:} We performed agreement analysis to identify image patches where all models consistently rated them as ``bad." It allowed us to summarize examples and patterns of these consensus ``failure" image patches, offering valuable insights for addressing nuclei segmentation challenges in kidney pathology.

\section{METHOD}
Fig.~\ref{fig:fig1} illustrates the overall framework of this study: (1) Large-scale diverse dataset creation, (2) Nuclei instance segmentation, (3) Rating on Segmentation Performance. Meaningful images were then identified through model agreement analysis. 

\subsection{Diverse Large-scale Dataset}
Our dataset incorporates both public and private kidney datasets, totaling 2,542 WSIs. According to Fig.~\ref{fig:fig1}b, 57\% (1,449 WSIs) are sourced from publicly available datasets such as Kidney Precision Medicine Project (KPMP)~\cite{kpmp_data}, NEPTUNE~\cite{barisoni2013digital}, and Human BioMolecular Atlas Program (HuBMAP)~\cite{hubmap-kidney-segmentation}, while 43\% (1,093 WSIs) are from an in-house collection at Vanderbilt University Medical Center. To enhance dataset diversity, we included WSIs from both human and rodent sources. As shown in Fig.~\ref{fig:fig1}c, the WSIs were stained using Hematoxylin and Eosin (H\&E), Periodic acid-Schiff methenamine (PASM), and Periodic acid-Schiff (PAS), with PAS being particularly prevalent in kidney pathology but less common in other organs. We randomly extracted four 512$\times$512-pixel image patches from each kidney WSI at 40$\times$ magnification. Patches that were contaminated, of low imaging quality, from broken tissue, or with impaired staining were discarded. This process resulted in an evaluation dataset of 8,818 image patches.
\subsection{Nuclei Instance Segmentation}
Nuclei instance segmentation was performed on these kidney nuclei image patches using three foundation models. We summarized the three foundation models and provided GitHub links to our customized GPU-optimized inference pipelines in Table.~\ref{table:models}: 
\begin{enumerate}
    \item \textbf{Cellpose}: Cellpose\cite{cellposs} is a generalist segmentation model that utilizes a U-Net backbone to predict the horizontal and vertical gradients of the topological maps, as well as a binary map. Through
a gradient tracking algorithm and grouping pixels that follows the same point (center), individual object can be retrieved.

    \item \textbf{StarDist}: StarDist\cite{stardist} uses a unique star-convex polygon representation for nuclei segmentation, which is particularly effective for roundish objects such as cell nuclei. A simple U-Net architecture is employed for dense prediction of radial distances and object probabilities, with non-maximum suppression used to select final instances.
   
    \item \textbf{CellViT}: CellViT\cite{cellvit} employs a hierarchical encoder-decoder Vision Transformer (ViT) backbone. The encoder utilizes pre-trained weights from a ViT trained on 104 million histological images (ViT\textsubscript{256})\cite{chen2022scaling}, addressing the data scarcity challenge often faced in medical imaging. The post-processing procedure follows the HoVer-Net methodology\cite{hover-net}.
\end{enumerate}
\begin{table}[h!]
\centering
\caption{SOTA cell nuclei Foundation models}
\begin{tabular}{p{2cm}cccc}
\toprule
Year-Month & Model & Backbone & Post-processing & Customized GPU Inference\\
\midrule
2022 Mar & StarDist (Histo.)\cite{stardist}& U-Net & Star-convex Polygon & \href{https://github.com/hrlblab/AFM_kidney_cells/tree/main/stage1_paper/stardist-inference-gpu}{Link URL} \\
\midrule
2022 Nov & Cellpose 2.0\cite{cellposs} & U-Net & GradientFlow Tracking & \href{https://github.com/hrlblab/AFM_kidney_cells/tree/main/stage1_paper/cellpose-inference-gpu}{Link URL} \\
\midrule
2023 Oct & CellViT\cite{cellvit} & ViT & HoVer-Net\cite{hover-net} & \href{https://github.com/hrlblab/AFM_kidney_cells/tree/main/stage1_paper/cellvit-inference-gpu}{Link URL} \\
\bottomrule
\end{tabular}
\label{table:models}
\end{table}

\subsection{Rating on Segmentation Performance}
Two pathologist-trained students scored the model predictions as “good,” “medium,” or “bad.” “Good” predictions correctly capture about 90\% of nuclei in an image patch, “bad” predictions capture less than 50\%, and the rest are categorized as “medium.” We apply this scoring system to quantitatively assess each model's predictions in our dataset. Finally, we ensemble ratings from three foundation models to curate strongly agreed-upon failure samples. These image patches can be used for training or fine-tuning future foundation models in kidney pathology.

\section{Experiments}

\textbf{Implementations}: The model inference pipeline was implemented using Python 3.9 and PyTorch 2.0.1, with GPU acceleration provided by CUDA 11.7. The experiments, conducted on an NVIDIA RTX A5000 GPU with 24GB memory, ensured efficient test-time inference. GPU-optimized inference codebases, are provided in the links in Table.~\ref{table:models}.

\subsection{Distribution of Rated Foundation Model Predictions}
We analyzed the distribution of predictions for each foundation model across our evaluation dataset. This analysis provides insights into the performance and behavior of each model by examining the frequency of different prediction classes (``good", ``medium", ``bad") within each foundation model. The prediction class assignments were determined through the manual rating-based curation.

\subsection{Agreement Analysis Between Foundation Models}
To further assess cross-model performance on our kidney nuclei dataset, we conducted an agreement analysis between the models. This analysis involved the following steps:
\begin{enumerate}
    \item \textbf{Agreement Matrix.} We calculated the agreement percentages between each pair of models to understand how often they make the same predictions. This matrix helps identify the consistency between foundation models. 
    \item \textbf{Class-Specific Agreement.} Image patches are categorized into three types: \textbf{All Three Models Agree}, which represents the number of patches where all models agree on the prediction class; \textbf{Two Models Agree}, indicating patches where any two models agree on the prediction class; and \textbf{No Agreement}, where none of the models agree on the prediction class.




\end{enumerate}
\subsection{Consensus ``Bad” Image Patches Mining}

By ensemble curated labels from three cell nuclei foundation models across our kidney dataset, we can identify strong-agreed-upon failure samples, which occur when all three models produce “bad” instance segmentation results. These “global failure” image patches, where all three models agree on poor performance, highlight the gap areas where the current cell nuclei foundation models need improvement in kidney pathology.
\section{RESULTS}

\subsection{Foundation Model Performance Evaluation}
\begin{figure*}[ht]
\begin{center}
\includegraphics[width=0.75\linewidth]{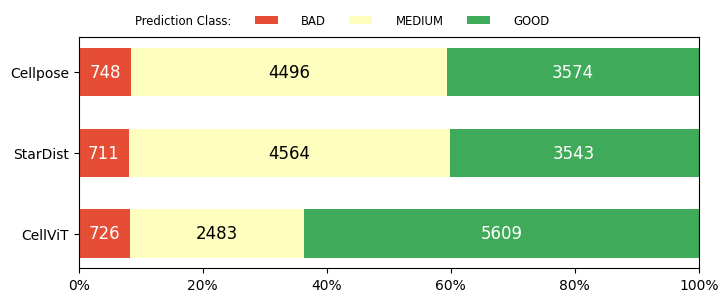}
\end{center}
\caption{The distribution of cell nuclei foundation model predictions across our dataset. Each row provides the number of ``good”, ``medium”, and ``bad” predictions made by the foundation model. }
\label{fig:in_model}
\end{figure*}

First of all, we rated the segmentation performance of these foundation models. Fig.~\ref{fig:in_model} shows the distribution of rated predictions. CellViT demonstrated superior performance in segmenting nuclei in these kidney images, with 5,609 (63.6\%) ``good" predictions, which outperforms Cellpose (40.5\%) and StarDist (40.2\%). This indicates that, despite not being specifically trained on kidney pathology nuclei data, the models retain some transferable knowledge applicable to our task. However, even with the best CellViT model, there is still  an existence of approximate 37\% of  rated predictions from ``medium" and ``bad," which underscores the need for a kidney domain-specific foundation model. 

\subsection{Foundation Model Agreement}
\begin{figure*}[ht]
\begin{center}
\includegraphics[width=1\linewidth]{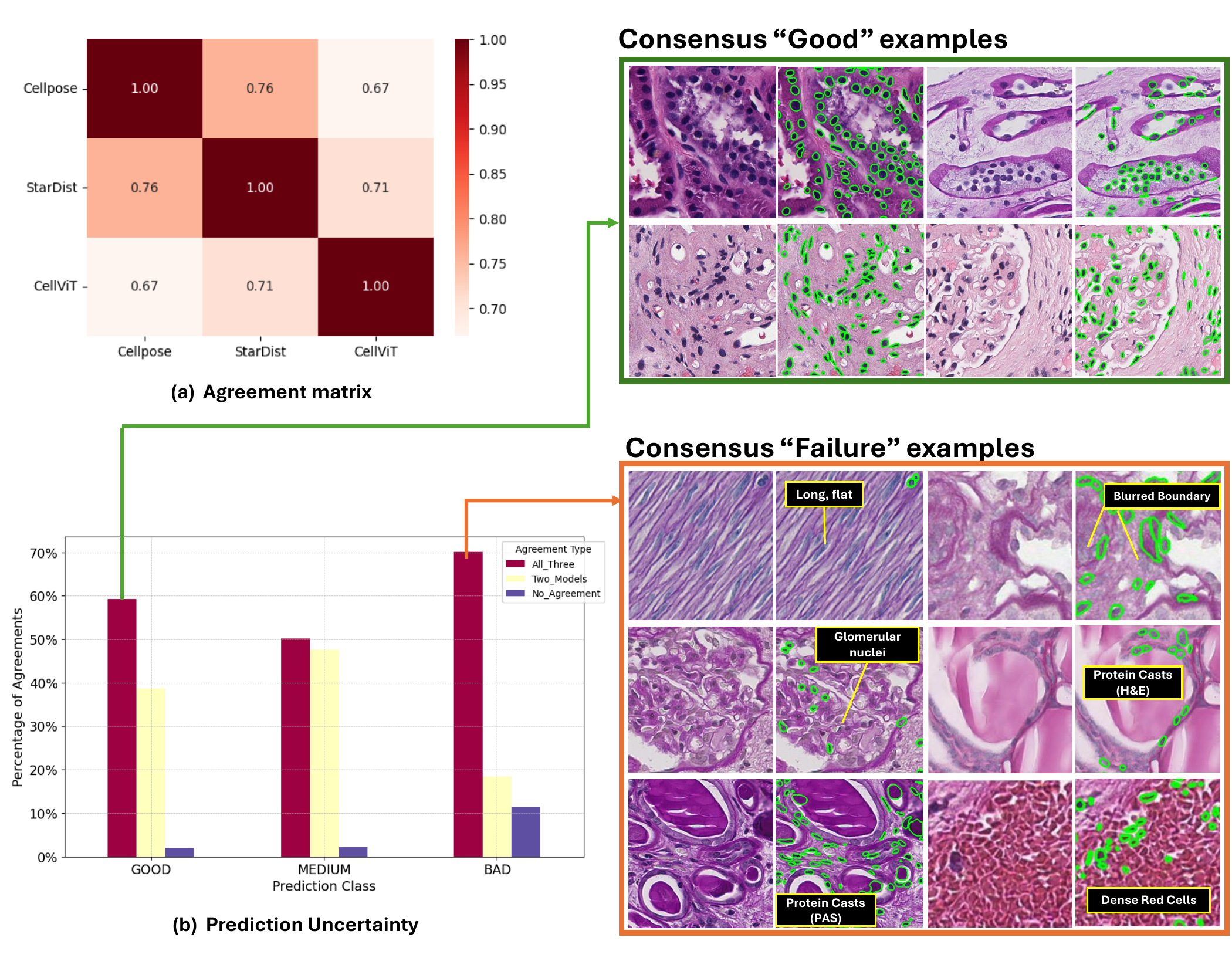}
\end{center}
\caption{(a) shows the agreement percentages between each pair of foundation models used in this study. To further assess the cross-model performance, (b) shows the percentages of image patches where \textbf{all three models agree}, \textbf{two models agree}, or \textbf{no models agree}, for each prediction class (``good”, ``medium”, ``bad”). Finally, the right panel presents examples of \textbf{Consensus “good”} and \textbf{Consensus “failure”} image patches, along with their respective \textbf{failure categories}.}
\label{fig:cross_models}
\end{figure*}

Fig.~\ref{fig:cross_models}a illustrates the consistency between model predictions. As shown in the Fig.~\ref{fig:cross_models}a, no pair exhibits over 90\% agreement or complete disagreement. The highest agreement is observed between Cellpose and StarDist (0.76), while the lowest is between Cellpose and CellViT (0.67). These non-extreme values, with a maximum of 0.76 and a minimum of 0.67, indicate that the current SOTA nuclei foundation models in digital pathology do not generalize exceptionally well to diverse large-scale kidney data. Each model has its own strengths, and this inconsistency inspires us to combine multiple SOTA foundation models to enhance performance in downstream kidney nuclei tasks. Both Fig.~\ref{fig:cross_models}b and Table.~\ref{tab:table2} show the number of image patches where all three models agree, two models agree, or no models agree for each prediction class (``good", ``medium", ``bad"). It is evident that ``bad" cases exhibit the highest inter-model reliability, with over 70\% agreement among all three models. 

\subsection{Ensemble Learning with Foundation Models}
Particularly, samples where all three foundation models agree in both the ``good" (\textbf{Consensus ``good"}) and ``bad" (\textbf{Consensus ``failure"}) classes are displayed, along with their respective \textbf{failure categories}. As shown, current nuclei foundation models perform well on image patches with darker nuclei and higher contrast between nuclei and non-nuclei regions. Conversely, Consensus ``failure" samples exhibit overall lower staining intensity, lighter nuclei, and reduced imaging contrast. Specifically, long and flat nuclei, nuclei with blurred boundaries, and densely distributed nuclei within glomeruli are particularly difficult to segment accurately. Additionally, slides that contain tubules filled with protein casts, or have an excessive number of red blood cells can lead to a higher incidence of false positive predictions.

\begin{table}[]
\centering
\vspace{0.4cm} 
    \caption{This table shows the percentages of model agreement in each rated prediction class. }


\resizebox{0.65\textwidth}{!}{%
\begin{tabular}{cccc}
\multicolumn{1}{l}{} &
  \cellcolor[HTML]{FFFFFF}{\color[HTML]{212121} \textbf{All\_Three}} &
  \cellcolor[HTML]{FFFFFF}{\color[HTML]{212121} \textbf{Two\_Models}} &
  \cellcolor[HTML]{FFFFFF}{\color[HTML]{212121} \textbf{No\_Agreement}} \\ \hline
\rowcolor[HTML]{F3F3F3} 
{\color[HTML]{212121} \textbf{GOOD}}   & {\color[HTML]{212121} 59.27} & {\color[HTML]{212121} 38.7}  & {\color[HTML]{212121} 2.03}  \\
\rowcolor[HTML]{FFFFFF} 
{\color[HTML]{212121} \textbf{MEDIUM}} & {\color[HTML]{212121} 50.23} & {\color[HTML]{212121} 47.57} & {\color[HTML]{212121} 2.2}   \\
\rowcolor[HTML]{F3F3F3} 
{\color[HTML]{212121} \textbf{BAD}}    & {\color[HTML]{212121} 70.17} & {\color[HTML]{212121} 18.4}  & {\color[HTML]{212121} 11.43}
\end{tabular}%
}
\label{tab:table2}
\end{table}

\section{Conclusion}
This study provides a comprehensive assessment of cell nuclei foundation models (Cellpose, StarDist, CellViT) in kidney pathology, focusing on nuclei instance segmentation. By evaluating a large and diverse kidney nuclei dataset, we identified CellViT as having superior performance. However, a performance gap remains between general and kidney-specific nuclei segmentation. Additionally, we utilized all three foundation models to curate consensus “good” and “failure” image patches, summarizing their characteristics to offer insights for improving cell nuclei foundation models in kidney pathology. An ongoing future direction will further mine these evaluation outcomes to support the training of kidney-domain-specific cell nuclei AI foundation models while minimizing annotation effort.

\acknowledgments 
This work was supported by the National Institutes of Health under award numbers R01EB017230, 1R01DK135597- 01, T32EB001628, and 5T32GM007347, and in part by the National Center for Research Resources and Grant UL1 RR024975-01. This study was also supported by the National Science Foundation (1452485, 1660816, and 1750213). The Vanderbilt Institute for Clinical and Translational Research (VICTR) is funded by the National Center for Advancing Translational Sciences (NCATS) Clinical Translational Science Award (CTSA) Program, Award Number 5UL1TR002243- 03. The content is solely the responsibility of the authors and does not necessarily represent the official views of the NIH or NSF. This work was also supported by Vanderbilt Seed Success Grant, Vanderbilt Discovery Grant, and VISE Seed Grant. We extend gratitude to NVIDIA for their support by means of the NVIDIA hardware grant. This works was also supported by NSF NAIRR Pilot Award NAIRR240055.
\bibliography{report} 
\bibliographystyle{spiebib} 

\end{document}